\documentclass[prl, aps, twocolumn, superscriptaddress, floatfix, nofootinbib, amsmath, amssymb, preprintnumbers]{revtex4-1}

\usepackage{graphicx}
\usepackage{dcolumn}
\usepackage{verbatim}
\usepackage{bm}
\usepackage{breakurl}
\usepackage{todonotes}
\usepackage{lmodern}

\usepackage{tikz}
\usepackage{pgfplots}
\pgfplotsset{compat=newest,every axis plot/.append style={line width=1pt}}
\usepackage[colorlinks,linktocpage,linkcolor=cyan,citecolor=cyan]{hyperref}

\usepackage{tabularx}

\definecolor{lightgray}{gray}{0.9}
\definecolor{Amber}{rgb}{1.0, 0.75, 0.0}
\definecolor{blizzardblue}{rgb}{0.67, 0.9, 0.93}

\begin{document}

\preprint{YITP-21-16}

\title{Beating the Lyth bound by parametric resonance during inflation}

\author{Yi-Fu Cai}
\email{yifucai@ustc.edu.cn}
\affiliation{Department of Astronomy, School of Physical Sciences, University of Science and Technology of China, Hefei, Anhui 230026, China}
\affiliation{School of Astronomy and Space Science, University of Science and Technology of China, Hefei, Anhui 230026, China}
\affiliation{CAS Key Laboratory for Researches in Galaxies and Cosmology, University of Science and Technology of China, Hefei, Anhui 230026, China}

\author{Jie Jiang}
\email{jiejiang@mail.ustc.edu.cn}
\affiliation{Department of Astronomy, School of Physical Sciences, University of Science and Technology of China, Hefei, Anhui 230026, China}
\affiliation{School of Astronomy and Space Science, University of Science and Technology of China, Hefei, Anhui 230026, China}
\affiliation{CAS Key Laboratory for Researches in Galaxies and Cosmology, University of Science and Technology of China, Hefei, Anhui 230026, China}

\author{Misao Sasaki}
\email{misao.sasaki@ipmu.jp}
\affiliation{Kavli Institute for the Physics and Mathematics of the Universe (WPI), UTIAS, The University of Tokyo, Chiba 277-8583, Japan}
\affiliation{Center for Gravitational Physics, Yukawa Institute for Theoretical Physics, Kyoto University, Kyoto 606-8502, Japan}
\affiliation{Leung Center for Cosmology and Particle Astrophysics, National Taiwan University, Taipei 10617}

\author{Valeri Vardanyan}
\email{valeri.vardanyan@ipmu.jp}
\affiliation{Kavli Institute for the Physics and Mathematics of the Universe (WPI), UTIAS, The University of Tokyo, Chiba 277-8583, Japan}

\author{Zihan Zhou}
\email{ustczzh@mail.ustc.edu.cn}
\affiliation{Department of Astronomy, School of Physical Sciences, University of Science and Technology of China, Hefei, Anhui 230026, China}
\affiliation{School of Astronomy and Space Science, University of Science and Technology of China, Hefei, Anhui 230026, China}
\affiliation{CAS Key Laboratory for Researches in Galaxies and Cosmology, University of Science and Technology of China, Hefei, Anhui 230026, China}


\begin{abstract}
We propose a novel mechanism for enhancing the primordial gravitational waves without significantly affecting the curvature perturbations produced during inflation. This is achieved due to non-linear sourcing of resonantly amplified scalar field fluctuations. Our result is an explicit scale-dependent counter-example of the famous Lyth bound, which opens up a promising perspective of producing detectable inflationary tensor modes with low-scale inflation and a sub-Planckian field excursion. We explicitly demonstrate the testability of our mechanism with upcoming Cosmic Microwave Background B-mode observations.
\end{abstract}


\maketitle


{\it Introduction --}
The breakthrough of direct gravitational wave (GW) detection by LIGO \cite{Abbott:2016blz} has heralded the birth of experimental GW cosmology. GWs carry invaluable information about various physical systems and production mechanisms, ranging from astrophysical mergers to the physics of the very early Universe. Among the vast variety of possible signals, the ones originating from quantum fluctuations of spacetime in the primordial Universe \cite{Grishchuk:1974ny, Starobinsky:1979ty} are of special interest as
their detection would provide information about physics in extremely high energy regimes,  and help unraveling the mystery of the origin of our Universe.  Due to their paramount physical significance many observational campaigns are aiming to detect them in near future (see \cite{Campeti:2020xwn} for a comprehensive overview). These include space-based GW detectors \cite{Bartolo:2016ami, Crowder:2005nr, Kawamura:2020pcg} and pulsar-timing surveys \cite{Kramer:2013kea, Burke-Spolaor:2018bvk}. Importantly, primordial GWs are expected to induce B-mode polarization on the Cosmic Microwave Background (CMB) \cite{Seljak:1996gy, Kamionkowski:1996ks}, and hence have been the focus of multiple CMB missions \cite{Aghanim:2018eyx, Ade:2018iql, Ade:2017uvt, Louis:2016ahn, Hanson:2013hsb, Dahal:2019xuf, Ade:2018sbj, Abazajian:2016yjj, Abazajian:2020dmr, Cai:2016hqj, Li:2017drr, Hazumi:2021yqq}.

As the prevailing paradigm of the primordial Universe, inflation \cite{Linde:2005ht,Liddle:2000cg} provides a compelling mechanism for generating primordial perturbations of both scalar and tensor types. Since the microscopic properties of inflation are still unknown, a significant effort is directed towards obtaining generic theoretical priors on the model space. Particularly, the observed nearly scale-invariant power spectrum of primordial curvature perturbations suggests that the inflaton potential is very flat, and the small amplitude of the observed CMB temperature fluctuations requires a hierarchy between a tower of slow-roll parameters, which are often related to certain mass hierarchies during inflation.
In the most standard models of inflation the amplitude of quantum tensor fluctuations is related to the inflationary energy scale.  The current upper bound of the tensor-to-scalar ratio $r$ suggests that inflation happens at relatively low energies compared to the Grand Unified Theory scale. Moreover, the tensor-to-scalar ratio sets a robust lower bound, known as the Lyth bound \cite{Lyth:1996im}, on the inflaton field excursion $ \Delta \phi$,
\begin{equation}\label{eq:Lyth}
 \Delta \phi \gtrsim \mathcal{O}(1) \Big( \frac{r}{0.01} \Big)^{1/2} M_\mathrm{Pl} ~. \nonumber
\end{equation}
The striking implication of this bound is that a measurement of $r$ at the level of its current upper bound would imply super-Planckian field excursions, which would pose a theoretical challenge for model building. Indeed, it is notoriously difficult to realize the delicate balance between the height and slope of the potential if higher-order operators also contribute significantly, which occurs when inflaton takes on a vacuum expectation value \cite{Copeland:1994vg, Dine:1995uk}. Thus, the effective field theory arguments favor a small-field variation, $\Delta\phi < M_\mathrm{Pl}$, unless certain symmetries are introduced to guarantee the absence of higher-order corrections. Additionally, it has been conjectured that potentials supporting large-field excursions might be incompatible with quantum gravity \cite{Ooguri:2006in}.
It is therefore pertinent to examine all the assumptions of the Lyth bound, and to propose compelling alternative scenarios where it is violated. This program becomes especially timely given the dramatic development of CMB B-mode experiments.

In this Letter, we put forward a novel mechanism for enhancing the primordial GWs compared to their production from vacuum fluctuations, hence beating the Lyth bound. Our mechanism relies upon the idea that amplified scalar field fluctuations can enhance GWs at higher orders in perturbation theory.  The scalar field fluctuations in our proposal are amplified due to a parametric resonance effect which is well-known in a broad range of physical systems, such as cosmic preheating \cite{Traschen:1990sw, Dolgov:1989us, Kofman:1994rk,Kofman:1997yn, Allahverdi:2010xz},  sound speed resonance of scalar fields \cite{Cai:2018tuh, Cai:2019jah, Chen:2019zza, Chen:2020uhe} and tensor modes \cite{Cai:2020ovp}, oscillating graviton mass \cite{Lin:2015nda, Kuroyanagi:2017kfx},  and extended axion-monodromy-like models \cite{Zhou:2020kkf, Chen:2010bka}.

With its minimalistic assumptions and flexible control over possible modifications in the scalar sector, our scenario stands aside from previously considered mechanisms. Early works in this direction have considered GW production due to direct coupling of additional scalar fields and inflaton \cite{Cook:2011hg, Senatore:2011sp, Biagetti:2013kwa, Ozsoy:2014sba,Mirbabayi:2014jqa}. These mechanisms generically alter the spectrum of scalar fluctuations along with the tensor spectrum. Another class of production mechanisms \cite{Dimastrogiovanni:2012ew, Maleknejad:2012fw, Adshead:2013qp, Dimastrogiovanni:2016fuu, Fujita:2017jwq, Ozsoy:2020ccy} relies on the presence of $U(1)$ or $SU(2)$ gauge sector and its coupling to an axion field, and hence, unlike our scenario, it predicts non-zero $EB$ and $TB$ cross CMB spectra \cite{Thorne:2017jft}.  Meanwhile,  the Lyth bound can be relaxed due to non-Bunch-Davies initial state in inflationary setting \cite{Ashoorioon:2013eia}.

{\it Our mechanism --} We consider two gravitationally interacting, canonical scalar fields $\phi$ and $\chi$. The role of the massless field $\chi$ is to generate the observed nearly-scale-invariant power spectrum for curvature perturbations, which are shown to be within the current observational bounds. The massive $\phi$ perturbations, on the other hand, are resonantly enhanced during inflation due to oscillatory mass term in the $\phi$ potential. These amplified scalar fluctuations source the GWs in a narrow band of scales, leading to strongly scale-dependent $r$.

On spatially-flat slicing, the Fourier modes of perturbations $\delta\chi$ and $\delta\phi$ evolve according to \cite{Sasaki:1995aw, Malik:2006ir, Arroja:2008yy}
\begin{align}
 &\ddot{\delta\chi}_k + 3H\dot{\delta\chi}_k + \frac{k^2}{a^2} \delta \chi _ k = \frac{\sqrt{2\epsilon_{\chi}}}{M_{\rm Pl}} \big[ \ddot{\phi}\delta\phi_k + \mathcal{S}_k \big],
\label{EoMQchi}\\
 &\ddot{\delta\phi}_k + 3H\dot{\delta\phi}_k + \left( \frac{k^2}{a^2} + \mathcal{M}_{\rm eff}^2 \right) \delta\phi_k = 0 ~,
\label{EoMQphi}
\end{align}
respectively, with the effective mass term $\mathcal{M}_{\rm eff}^2$ being determined by the background dynamics to be specified later and $\epsilon_\chi$ denoting the first slow-roll parameter.
The right-hand side of Eq.~\eqref{EoMQchi} contains both the first and second order contributions, where the latter is given by \cite{Malik:2006ir, Arroja:2008yy}
\begin{align}\label{eq:scalar-source}
 \mathcal{S}_k = \int \frac{ d^3 \bm{ p } } { ( 2 \pi ) ^ 3 }
 \Big\{ & \frac{\bm{p}\cdot\bm{k}}{k^2} \left[ \frac{ ( \bm{ p } - \bm{ k } ) ^ 2 } { a ^ 2 } + \mathcal{M}_{\rm eff}^2 \right] \nonumber\\
 & -\frac{\bm{p} \cdot (\bm{k} - \bm{p})}{ 2 a ^ 2 } \Big\} \delta\phi_{\vert\bm{p}\vert} \delta\phi_{\vert\bm{k} - \bm{p}\vert} ~.
\end{align}

Our goal is to construct $\mathcal{M}_{\rm eff}^2$ such that $\delta\phi$ modes exhibit significant amplification with respect to their vacuum values.  At the same time we should make sure that $\delta\chi$ perturbations are not significantly affected because these determine the tightly constrained spectrum of curvature perturbations.

The amplified $\delta\phi$ modes source GWs at second order in perturbation theory. Indeed,  the transverse-traceless tensor modes $h^\lambda_{k}$ propagate according to
\begin{equation}\label{FouierEoM}
 \ddot{h}^\lambda_{k} + 3 H\dot{h}^\lambda_{k} + \frac{k^2}{a^2}h^\lambda_{k}
 = \mathcal{T}^\lambda_{k}(t) ~,
\end{equation}
where non-linear source term $ \mathcal{ T } _ { k } ^ \lambda(t) $ takes the form
\begin{align}\label{eq:tensor-source}
 \mathcal{T}_{k}^{\lambda}(t) =&
 \frac{2}{M_p^2} \int \frac{d^3{\bm p}}{(2\pi)^{3}} e^{ \lambda } _ {ij} ( {\bm k} ) \frac{p_i p_j}{a^2} \delta\phi_{\vert{\bm p}\vert} \delta\phi_{\vert{\bm k}-{\bm p}\vert} ~,
\end{align}
with $e^{\lambda}_{ij}({\bm k})$ being the polarization tensor.  Scalar-induced GWs are widely studied in numerous contexts (see e.g. \cite{Matarrese:1993zf, Ananda:2006af, Baumann:2007zm, Saito:2008jc, Cai:2018dig, Braglia:2020taf}).

We should impose three main conditions on $\mathcal{M}_{\rm eff}^2$.  First of all, it should exhibit oscillatory features during the sub-Hubble evolution of certain modes of interest. Due to these oscillatory terms in Eq.~\eqref{EoMQphi}, the fluctuations $\delta\phi$ experience a Mathieu-type parametric resonance growth in a narrow band around the resonant scale. These amplified modes will then induce enhanced GWs.  Let us comment here that the parametric resonance is not the only possible amplification mechanism. Tachyonic instability, for example, might lead to similar enhancement. We however do not study this possibility here and concentrate on the model featuring parametric resonance.

As a second condition on $\mathcal{M}_{\rm eff}^2$ we require it to become small right after horizon crossing.  From Eq.~\eqref{eq:scalar-source} it is clear that $\mathcal{M}_{\rm eff}$ contributes to the non-linear backreaction to the $\delta \chi$ sector. Generically, this contribution is non-negligible during several $e$-folds after Hubble crossing, and if $\mathcal{M}_{\rm eff}$ is large enough during this period,  the enhanced $\delta\phi$ modes would spoil the successful realization of the quasi-scale-invariant curvature power spectrum.  By making  $\mathcal{M}_{\rm eff}^2$ small during this epoch suppresses the non-linear contribution of $\delta\phi$ on $\delta\chi$, and, importantly, allows $\delta\phi$ not to immediately decay after horizon crossing,  therefore giving the gravitational waves more time to be enhanced by the amplified $\delta\phi$ modes.  Later we will see that a simple linear potential for $\phi$ during this epoch is enough for sufficiently suppressing the backreaction from $\delta \phi$.

The third requirement on the mass term is that it should eventually become large enough, $\mathcal{M}_{\rm eff}^2\gtrsim \mathcal{O}(H^2)$, to stop the evolution of $\phi$. This crucial requirement is necessary for forbidding $\delta\phi$ affecting the curvature power spectrum directly. This can be easily understood with the help of the $\delta N$ formalism, a powerful tool for computing the power spectrum of primordial curvature perturbations $P_\mathcal{R}(k)$ \cite{Starobinsky:1982ee, Starobinsky:1986fxa, Sasaki:1995aw, Lyth:2004gb, Abolhasani:2019cqw}. On super-Hubble scales the final value of the conserved curvature perturbation $\mathcal{R}_c$ can be expressed in terms of the variation in number of $e$-folds $\delta N$ computed back from the adiabatic limit, $\mathcal{R}_c(\boldsymbol{x})=\delta N(\boldsymbol{x},t)$. Stopping the evolution of $\phi$ allows the number of e-folds, hence also the resulting curvature spectrum to be governed predominantly by $\chi$.

{\it Concrete example --}
Inspired by the axion-monodromy inflation \cite{Silverstein:2008sg, McAllister:2008hb} and relaxion mechanism \cite{Graham:2015cka, Espinosa:2015eda}, we consider a potential with oscillatory modulations. We particularly consider the following concrete example of two-field inflation (see \cite{Cai:2019bmk,Zhou:2020kkf} for related models in context of primordial black hole formation)
\begin{align}\label{Potentialeq}
 &V( \phi, \chi ) =  V_0 \Big( 1 - \sqrt{2 \epsilon_\phi}\frac{\phi}{M_\mathrm{Pl}} + \sqrt{2 \epsilon_\chi}\frac{\chi}{M_\mathrm{Pl}} + \eta_\chi \frac{\chi^2}{2M_\mathrm{Pl}^2} \Big) \nonumber \\
 & + \frac{1}{2}\Lambda^4(\phi)\cos(\frac{\phi}{f_\mathrm{a}}) \Big[ \tanh\big(\frac{\phi_\mathrm{s} - \phi}{f_\mathrm{a}}\big) + 1 \Big] + V_\mathrm{m}(\phi)~.
\end{align}
Here, $V_0$ determines the inflationary energy scale, while $\epsilon_\chi$ and $\eta_\chi$ determine the spectrum of curvature perturbations. The crucial ingredient is the oscillatory feature in the $\phi$-sector. Particularly, the mass scale $f_a$, playing the role of the axion decay constant, determines the period of oscillations, while the function $\Lambda(\phi)$, chosen to be $\Lambda(\phi) = \Lambda_0 \big( 1 + \phi/M \big)$ with a constant scale $M$, determines the amplitude of oscillation barriers.

\begin{figure}[htp!]
    \centering
    \includegraphics[width=\columnwidth]{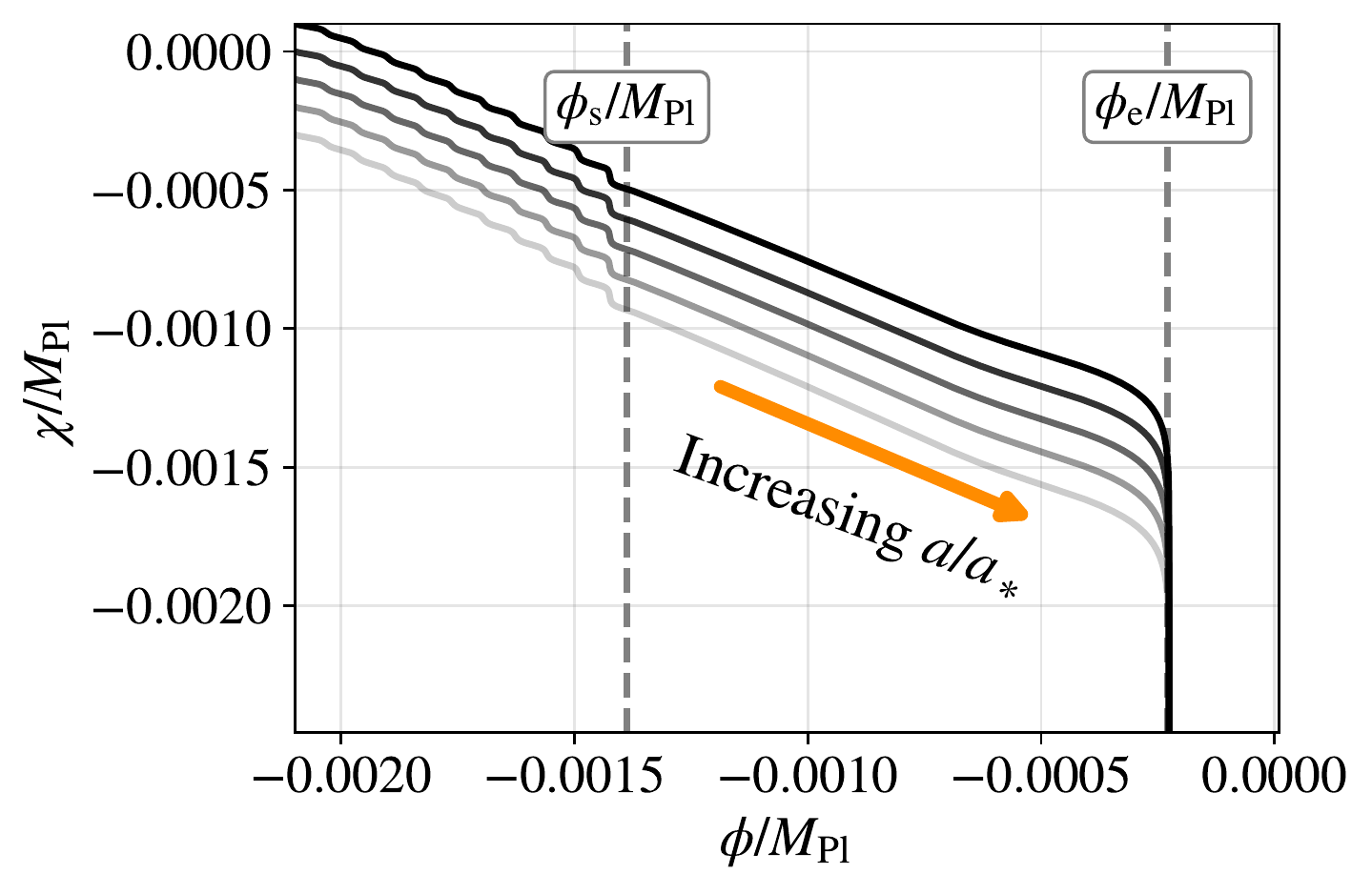}
    \includegraphics[width=\columnwidth]{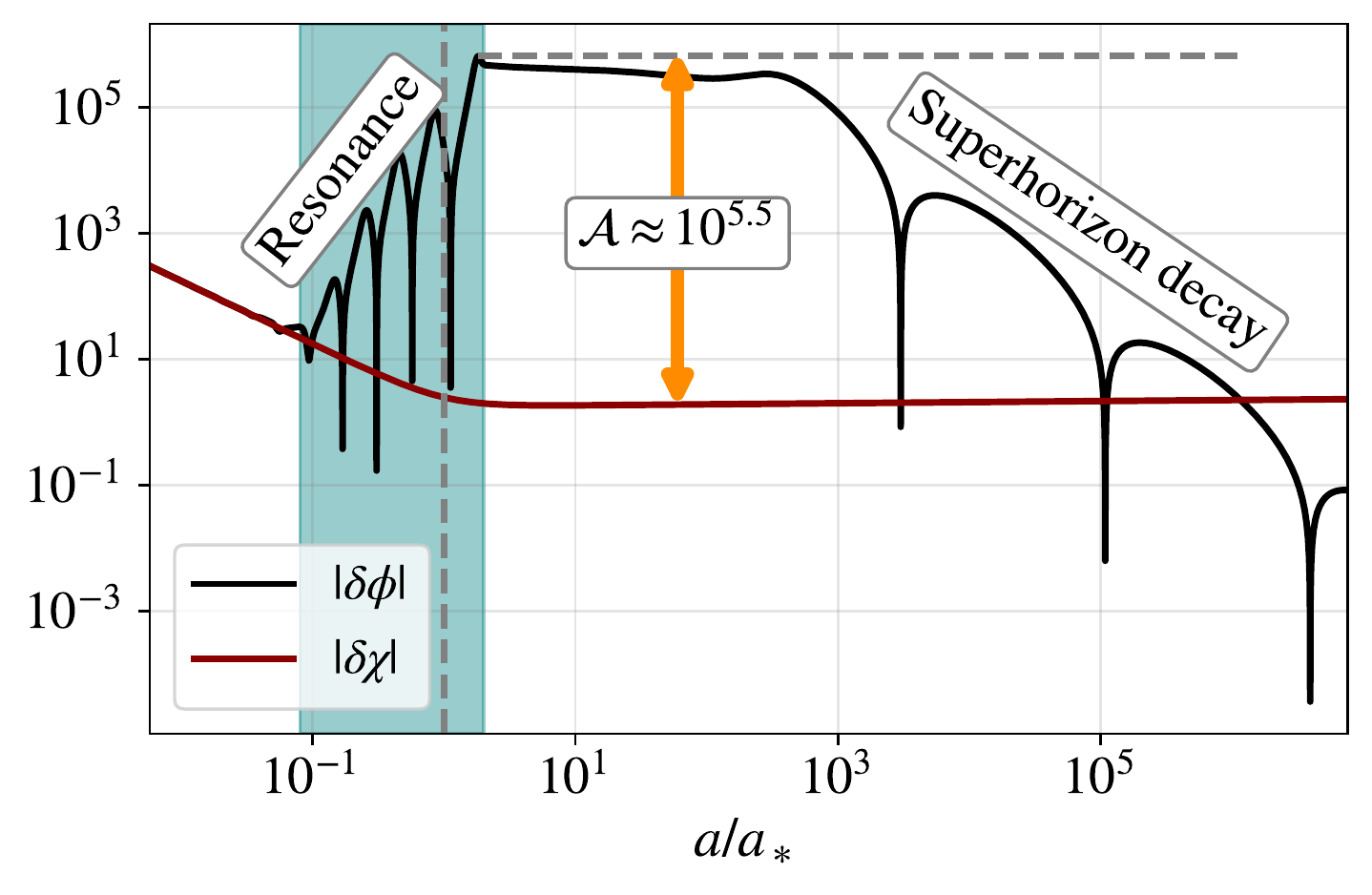}
    \caption{
    \textit{Upper panel:}
    Evolution of background trajectories in the ($\phi, \chi$) field space for various initial conditions.
    The field excursions are $\Delta\phi \sim \Delta\chi \sim 0.001 M_\mathrm{Pl}$, hence maintaining small-field inflationary regime.
    \textit{Lower panel:}
    Evolution of the perturbations $\delta\phi$ (in black) and $\delta\chi$ (in red) as a function of the scale factor $a$,  for a mode $k_\ast$ in the resonant band.  $\delta\phi$ experiences a resonant amplification when $k_\ast$ is sub-Hubble,  freezes for a short period after $k_\ast$ crosses the horizon at $a_\ast$,  and decays after $\phi$ is stabilized at $\phi_\mathrm{e}$.
    $\delta\chi$, on the other hand, is damped until Hubble-radius exit and freezes afterwards.
    As such, $\delta \phi$ is resonantly amplified by a factor $\mathcal{A} \sim 10^{5.5}$ with respect to its (non-amplified) vacuum value.
    The parameters are chosen to be: $\epsilon_\phi = 1.4\times 10^{-8}$, $\Lambda_0=2.4\times 10^{-6}M_\mathrm{Pl}$, $M=0.01 M_\mathrm{Pl}$, $f_a=1.0\times 10^{-5}M_\mathrm{Pl}$,  $\phi_\mathrm{s} = -1.4\times 10^{-3}M_\mathrm{Pl}$, $V_0=3.0\times 10^{-15}M_\mathrm{Pl}^4$, $\epsilon_\chi=5.7\times 10^{-9}$ and $\eta_\chi=-0.016$. }
    \label{fig:resonance}
\end{figure}

During the initial phase the field $\phi$ probes the oscillatory segment of the potential. The effective mass is $\mathcal{M}_{\rm eff}^2 \simeq -\Lambda^4(\phi) {\rm cos}(\phi/f_a) / f_a^2$ and therefore our first condition on $\mathcal{M}_{\rm eff}^2$ is satisfied,  leading to the parametric resonance.  At around Hubble-radius crossing of the resonant mode the potential is succeeded by a non-oscillatory linear segment, during which the enhanced $\delta\phi$ modes stay approximately constant (the second requirement on $\mathcal{M}_{\rm eff}^2$).  Finally,  this stage is stopped by the mass term $V_\mathrm{m}(\phi)$ at around $\phi_\mathrm{e}$,  after which the only dynamical evolution is due to $\chi$ (the third requirement on $\mathcal{M}_{\rm eff}^2$). Due to the plateau potential $V_0$ and small-field inflation, the number of e-folds is mainly governed by $\chi$ while the contribution from $\phi$ evolution is negligible,  $N = {\chi^2}/{(2M_\mathrm{Pl}^2)} + {\chi}/{(\sqrt{2 \epsilon_\chi}M_\mathrm{Pl})} + \cdots.$ The resulting power spectrum $P_\mathcal{R}(k)$ is therefore fully dominated by $\delta\chi_k$.  The background dynamics is depicted in the upper panel of Fig.~\ref{fig:resonance}, where different lines correspond to different initial $\chi$-values -- all converging to the same final trajectory as the $\phi$-field stops evolving.  The field excursions $\Delta\phi$ and $\Delta\chi$ are both limited, ensuring that we remain in the regime of small-field slow-roll inflation.

{\it Results --} During the oscillatory stage Eq.~\eqref{EoMQphi} can be approximately recast into a form of Mathieu equation.  According to the Floquet theory the amplification of $\delta\phi_{k_\ast}$ in a resonant band can be captured by $\vert\delta\phi_{k_\ast}\vert \propto e^{\lambda_{k_\ast} Ht}$ with $\lambda_{k_\ast} = \mu_{k_\ast} M_\mathrm{Pl}\sqrt{2\epsilon_\phi}/2 f_a - \frac{3}{2}$,
where $\mu_{k_\ast}$ is the Floquet exponent. The amplification factor $\mathcal{A}\equiv |\delta\phi_{k_*}| / |\delta\phi_{k_\ast,vac}|$ measures the relative amplification with respect to the non-amplified vacuum result $\delta\phi_{k_\ast,vac}$.

In order for our proposal to operate successfully, we need to check two main conditions.  First of all, we need our parameter space to allow for large enough $\lambda_{k_\ast}$ in order to get a significant amplification factor $\mathcal{A}$. Second, we need to achieve a nearly scale-invariant power spectrum $P_\mathcal{R}=A_s(k/k_p)^{n_s-1}$, where $A_s=H^2/8\pi^2 M_\mathrm{Pl}^2 \epsilon_\chi$ is the amplitude and $n_s\simeq1+2\eta_\chi$ is the spectral index at the pivot scale $k_p\simeq 0.05 \text{Mpc}^{-1}$. This requires $\delta\chi$ not to be dominated by the linear and non-linear backreaction from the amplified $\delta\phi$ sector.  The requirement of small linear backreaction is easy to quantify and it can be formulated as a simple condition $\mathcal{F}_\mathrm{L} \equiv \mathcal{A} \vert\ddot\phi /M_\mathrm{Pl}\vert \sqrt{2\epsilon_\chi}/H^2 \approx 6\mathcal{A} \sqrt{\epsilon_\phi\epsilon_\chi} \ll 1$. To verify the validity of this condition let us consider a simple rescaling of our model parameters: $\epsilon_\chi \rightarrow R^{2p}\epsilon_\chi, \epsilon_\phi \rightarrow R^{2q}\epsilon_\phi,  H^2 \rightarrow R^{2p}H^2,  f_\mathrm{a} \rightarrow R^{q}f_\mathrm{a}$,  where $R,  p$ and $q$ are constants.  Notice that both $\mathcal{A}$ and $A_s$ are invariant under these rescalings,  while $\mathcal{F}_\mathrm{L}$ transforms as $\mathcal{F}_\mathrm{L} \rightarrow R^{p + q}\mathcal{F}_\mathrm{L}$. This guarantees that by appropriately choosing the $R,  p$ and $q$ factors we can sustain a large amplification factor $\mathcal{A}$, while suppressing the transfer coefficient $\mathcal{F}_\mathrm{L}$.
We will shortly demonstrate numerically that in this regime the non-linear effects on the scalar power spectrum are relatively small. In other words, we have access to a large parameter space sustaining both a large amplification factor, and a nearly scale-invariant curvature power spectrum.

By initiating the mode functions in the Bunch-Davies vacuum state, $a\delta\phi_k(\tau_0) = a\delta\chi_k(\tau_0) = e^{-ik\tau}/\sqrt{2k}$, we numerically solve Eqs.~\eqref{EoMQchi} and \eqref{EoMQphi}. The results for a particular set of model parameters are shown in the lower panel of Fig.~\ref{fig:resonance}. We can clearly see the resonant amplification of $\delta\phi_{k_\ast}$, its freezing after Hubble-radius crossing, and the subsequent decay.

\begin{figure}[htp!]
    \centering
    \includegraphics[width=\columnwidth]{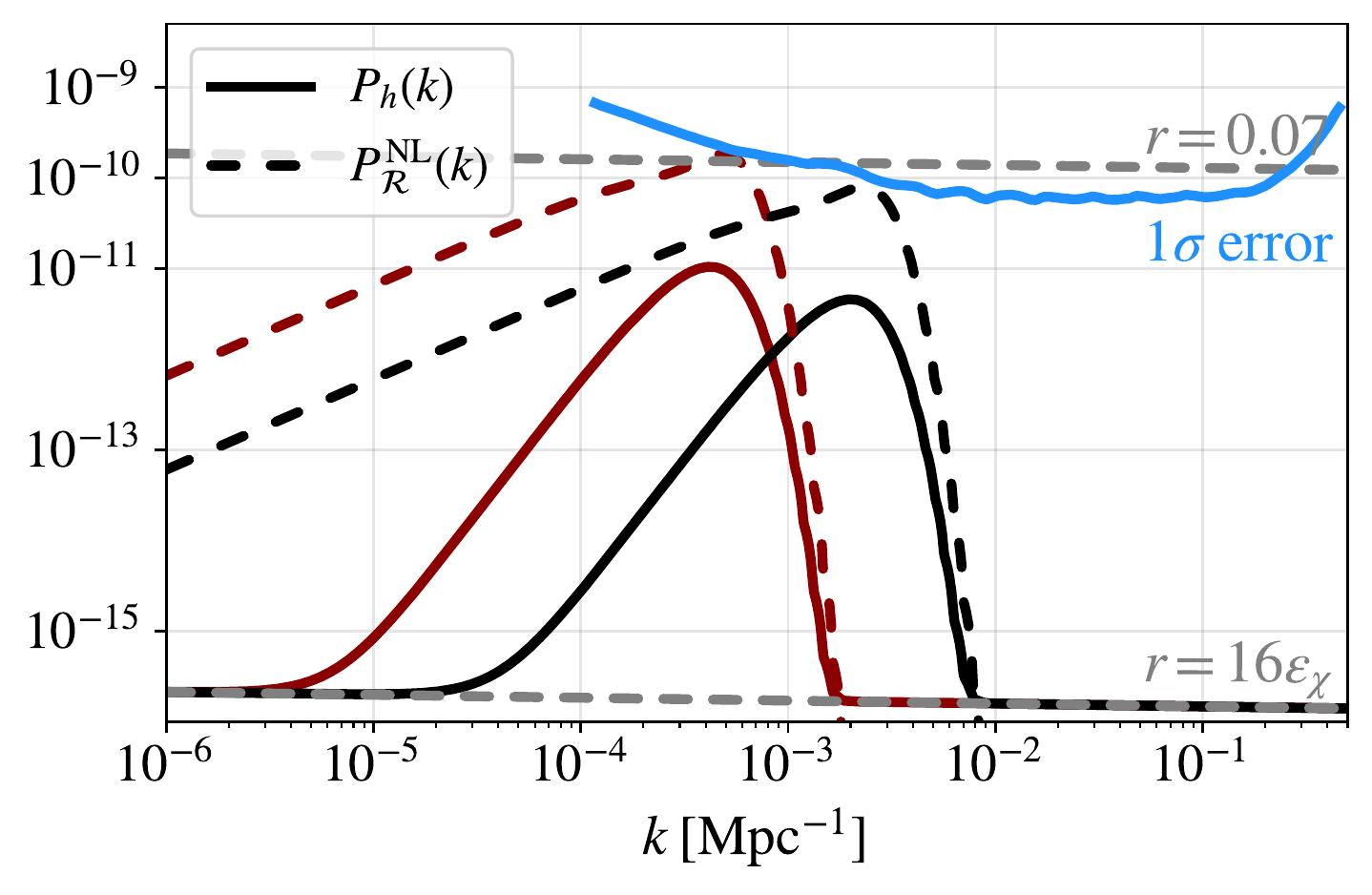}
    \includegraphics[width=\columnwidth]{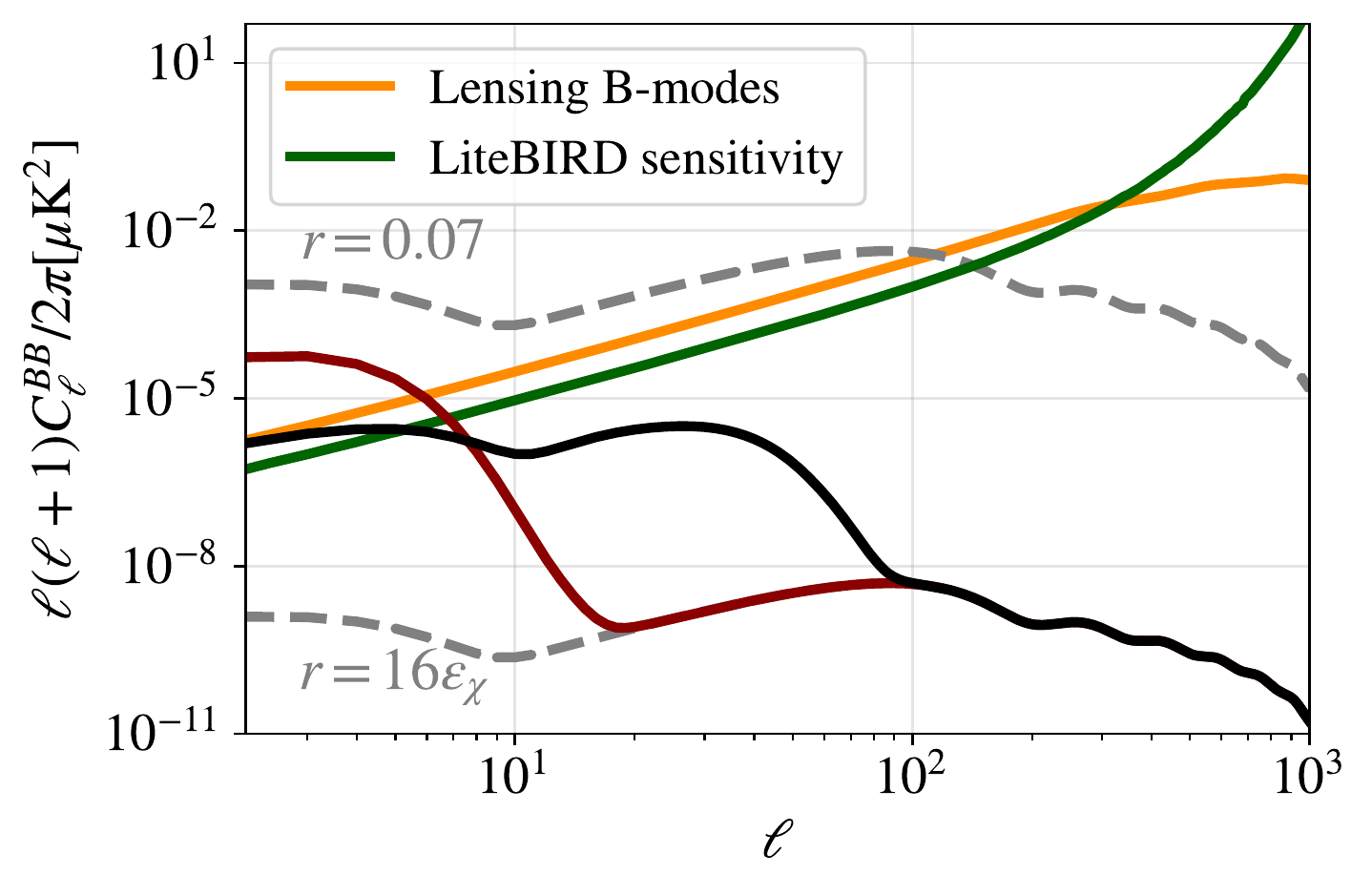}
    \caption{
    \textit{Upper panel:} GW power spectra induced by resonantly enhanced $\delta\phi$ fluctuations during inflation, depicted for two different resonance modes $k_\ast \sim 10^{-3.5}$ and  $10^{-2.5} \mathrm{Mpc}^{-1}$ (in red and black solid lines respectively), together with the corresponding non-linear corrections to the curvature power spectrum (dashed lines).  Also shown are the nearly scale-invariant tensor spectra with $r=16\epsilon_\chi$ and $r=0.07$ (gray, dashed lines),  and the $1-\sigma$ upper limit on the curvature power spectrum digitized from Ref.~\cite{Akrami:2018odb}.
    \textit{Lower panel:} Corresponding unlensed B-mode CMB angular power spectra with best-fit cosmological parameters from Ref.~\cite{Aghanim:2018eyx}.  For comparison, we also show the expected sensitivity curve of the LiteBIRD mission~\cite{Ishino:2016izb} (in green), the lensing B-mode spectrum (in orange), and the predictions for nearly scale-invariant tensor spectra (in gray).
    }
    \label{fig:powerspectrum}
\end{figure}

After finding the amplified scalar mode functions we have numerically obtained the corresponding primordial tensor power spectrum $P_h$ shown in the upper panel of Fig.~\ref{fig:powerspectrum} for two different resonant scales (red and black solid lines).
We clearly see that GWs are significantly amplified with respect to the vacuum contribution $P_h^{\rm vac}(k) = 2 H^2 / \pi^2 M_\mathrm{Pl}^2$ (corresponding to $r^{\rm vac}=16\epsilon_\chi$), while their peaks are within the current upper limit of $r\lesssim 0.07$.  The spectra have an infrared scaling of $\sim k^3$,  while in the ultraviolet regime they are approximated as $ P_h^{\rm ind}(k) \sim \exp\left[-\ln^2 (k/k_\ast) /2\Delta^2\right]$ (see also \cite{Cai:2019cdl, Pi:2020otn}).

Finally, we calculate the associated non-linear corrections to the scalar power spectrum, $P^\mathrm{NL}_\mathcal{R}$. This is shown as red and black dashed curves in the upper panel of Fig.~\ref{fig:powerspectrum}. We find that those second-order corrections do not dominate over the standard linear power spectrum. Thus, the tight observational constraints on the shape and amplitude of the power spectrum remain respected by the present mechanism.

In order to assess the observational implications of our scenario, we also compute the CMB B-mode polarization spectra.  The lower panel of Fig.~\ref{fig:powerspectrum} shows the primordial (unlensed) B-mode spectra obtained with the \href{http://camb.info}{\texttt{CAMB}} code \cite{Lewis:1999bs}. For reference we also show the expected LiteBIRD sensitivity curve \cite{Ishino:2016izb}, as well as the CMB lensing by the cosmic large-scale structure.  It can be seen that our scenario predicts a clear and unique signal of B-mode polarization which is distinguishable from the standard prediction of single-field slow-roll models. Generically,  at large angular scales the resonantly-induced signal dominates over the vacuum contribution, while on smaller scales the latter takes over.  We find that for $k_\ast \lesssim 10^{-3.5} \text{Mpc}^{-1}$ we have a significant enhancement of the reionization peak.  For $k_\ast \sim 10^{-2.5} \text{Mpc}^{-1}$ this enhancement is weaker but is still above the sensitivity limit.  It is interesting to note that in this case there is also a second peak at $\ell \sim 30$.  While this is below the sensitivity curve shown in the figure,  it would be observable with further experimental improvements, and such double-peaked structure would provide a smoking-gun evidence for our scenario.  In general, when primordial GWs are measured,  a significant effort should be directed towards checking their consistency with the vacuum result.  In our scenario, where GWs are produced by matter sources, the tensor spectrum strongly deviates from scale-invariant shape,  and any deviation from scale-invariant spectrum can be considered as a strong hint for presence of GW sources during inflation.  The two major classes of tensor enhancement mechanisms, namely the ones in spirit similar to our proposal, and the gauge field models discussed earlier, can be distinguished from each other by measuring the chirality of the primordial GW background. Results reported here serve as an additional motivation for a full-sky B-mode survey, as well as developing robust CMB delensing techniques.

{\it Concluding remarks --} In this Letter we propose an innovative possibility for generating enhanced primordial GWs in the scope of slow-roll inflation while the observed power spectrum of curvature perturbations remains within observational bounds.  Such a `{\it mission impossible}' becomes possible due to non-linearly induced GWs by scalar field fluctuations at sub-Hubble regime undergoing a parametric resonance. Notably, our mechanism provides a scale-dependent counter-example of the Lyth bound within slow-roll inflation. Our scenario can generate detectable primordial GWs for the forthcoming CMB polarization experiments even within inflationary models with very low energy scales and limited excursion range of the inflaton field. In order to illustrate the testability of this mechanism we calculate the CMB B-mode spectra and compare them with the expected sensitivity of the upcoming LiteBIRD mission.

The proposed mechanism would lead to several important implications which deserve careful follow-up studies. From the theoretical standpoint, our mechanism resolves the challenge of constructing models supporting sub-Planckian field excursions, even if future observational surveys detect large primordial GWs.  It is interesting to mention that our mechanism breaks the traditional scaling law relating the tensor-to-scalar ratio of inflation to the field range presented in \cite{Mirbabayi:2014jqa}. This is because while in such a scaling law the enhanced tensor and curvature perturbations are sourced from the same sector, in our scenario they are generated by two different, only gravitationally coupled sectors. As a result we have $r\sim \epsilon^0$ instead of $r \sim \epsilon ^2$ in \cite{Mirbabayi:2014jqa}.  The discovery of such a new scaling relation in our mechanism should be comprehensively investigated in the near future.

An important question is whether the standard Lyth bound can be generalized to accommodate the possibility of resonance effects present in our proposal. From phenomenological perspective our mechanism is able to produce enhanced primordial GWs without introducing any exotic matter fields or modified gravity effects. Nevertheless, the forms of the second order scalar and tensor source terms in Eqs.~(\ref{eq:scalar-source}) and (\ref{eq:tensor-source}), respectively, forced us to construct a potential with several special features in order to induce the resonant amplification. This is because we have focused on minimal coupling of the resonant scalar field $\phi$ to gravity. It would be interesting to extend the current idea to scenarios with non-minimal couplings to gravity. In general, developing an extended, model-independent framework and studying the implications for induced GWs is an interesting direction to pursue.  Moreover,  in a follow-up work it would be desirable to construct a model for our mechanism from the first principles of fundamental physics laws.

Finally,  while the scalar perturbations produced in our model are essentially indistinguishable from standard inflationary predictions within current experimental sensitivities, future improvements, especially on larger scales, can potentially lead to detection of the additional contributions to the scalar spectrum; see the dashed curves on the upper panel of  Fig.~\ref{fig:powerspectrum}.  Moreover, the non-linear interactions with $\delta\phi$ would inevitably render $\delta\chi$ perturbations more non-Gaussian compared to their single-field counterparts. As a result, the standard consistency condition would be violated in our scenario.  We therefore anticipate that the parametric resonance would imprint potentially observable non-Gaussian effects non-linearly on both the curvature perturbations and primordial GWs with unique scale-dependent shapes.  These exciting possibilities should be studied in detail,  and we leave their careful exploration to future work.

{\it Acknowledgments --}
We are grateful to Yashar Akrami, Masashi Hazumi, Eiichiro Komatsu, Andrei Linde, Karim Malik, Shi Pi, Alexei Starobinsky, Dong-Gang Wang, Yi Wang and Pierre Zhang for very valuable comments on this article.
YFC, JJ and ZZ are supported in part by the National Key R$\&$D Program of China (2021YFC2203100),  by NSFC (11653002, 11961131007, 11722327, 11421303), by the National Youth Talents Program of China, by the Fundamental Research Funds for Central Universities, by the CSC Innovation Talent Funds, by CAS project for young scientists in basic research (YSBR-006),  and by the USTC Fellowship for International Cooperation.
MS is supported in part by JSPS KAKENHI grants (19H01895, 20H04727, 20H05853).
VV is supported by the WPI Research Center Initiative, MEXT, Japan, and by a JSPS KAKENHI grant (20K22348). We acknowledge the use of computing facilities of Kavli IPMU, as well as the clusters {\it LINDA} and {\it JUDY} of the particle cosmology group at USTC.

\bibliography{bibliography_resonant_GW}

\end{document}